\documentclass{article}
\usepackage{kr} 

\usepackage{times}
\usepackage{soul}
\usepackage{url}
\usepackage[hidelinks]{hyperref}
\usepackage[utf8]{inputenc}
\usepackage[small]{caption}
\usepackage{graphicx}
\usepackage{amsmath}
\usepackage{amsthm}
\usepackage{booktabs}
\usepackage{algorithm}
\usepackage{algorithmic}
\usepackage{tikz}
\usepackage{subcaption}
\usepackage{xspace}

\newcommand{\norm}[1]{\left\|#1\right\|}
\newcommand{\form}{\Phi}
\newcommand{\ourtool}{$\mathtt{d4}\text{-}\mathtt{dyn}$\xspace}

\newcommand{\ganak}{{\tt ganak} }

\newtheorem{example}{Example}

\title{Efficient Incremental \#SAT via Cross-Instance Knowledge Reuse}

\author{%
Uriya Bartal$^1$\and
Dror Fried$^1$\and
Jean-Marie Lagniez$^2$ \\
\affiliations{}
$^1$The Open University Of Israel, Raanana, Israel\\
$^2$CRIL, U. Artois \& CNRS, F-62300 Lens, France \\
\emails{}
bauriya2@post.openu.ac.il,
dfried@openu.ac.il,
lagniez@cril.fr
}
\begin{document}

\maketitle

\begin{abstract}
Model counting ($\#\text{SAT}$) is a fundamental yet $\#\text{P}$-complete problem central to probabilistic reasoning. 
In this work, we address \textit{incremental model counting}, where sequences of structurally similar formulas must be counted. 
We propose an approach that amortizes computation via a persistent caching mechanism, retaining component data across solver calls to avoid redundant search. Additionally, we investigate branching heuristics adapted for this setting. We focus on the problems of argumentation and soft core, for which incremental model counting is natural.
Experiments demonstrate that our method improves performance compared to current model counters, highlighting the capability of structure-aware reuse in dynamic environments.
\end{abstract}

\section{Introduction}\label{Sec:intro}

Model counting (also known as \#SAT) is a fundamental problem in constraint programming and automated reasoning, with key applications in artificial intelligence~\cite{DBLP:conf/aips/PalaciosBDG05,DBLP:conf/aips/DomshlakH06,DBLP:journals/ijar/IzzaHINCM23}, product configuration~\cite{DBLP:conf/ictai/AstesanaCF10,DBLP:journals/tosem/SundermannRHTS24}, and other domains. 
The goal is to compute or estimate the number of solutions that satisfy a given set of Boolean constraints. 
Model counting is an inherently challenging task~\cite{DBLP:journals/siamcomp/Valiant79}, and thus presents significant scalability challenges~\cite{DBLP:journals/jea/FichteHH21}. 
Nevertheless, the central role of model counting has motivated substantial research into the development of both exact and approximate model counting algorithms capable of handling increasingly large and complex instances. 
This ongoing progress is exemplified by recent model counting competitions~\cite{DBLP:journals/corr/abs-2504-13842} that demonstrate the advancement of tools and provide rigorous benchmarks for state-of-the-art solvers.
A wide range of (exact) model counters have been introduced and investigated, including search-based approaches such as {\tt cachet}~\cite{DBLP:conf/sat/SangBBKP04}, {\tt SharpSAT}~\cite{DBLP:conf/sat/Thurley06}, {\tt SharpSAT-TD}~\cite{DBLP:conf/cp/KorhonenJ21}, and \ganak~\cite{DBLP:conf/cav/SoosM25}, as well as compilation-based approaches like {\tt C2D}~\cite{DBLP:conf/ecai/Darwiche04}, {\tt SDD}~\cite{DBLP:conf/ijcai/OztokD15}, {\tt dSharp}~\cite{DBLP:conf/ai/MuiseMBH12}, {\tt eadt}~\cite{DBLP:conf/ijcai/KoricheLMT13}, and {\tt d4}~\cite{DBLP:conf/ijcai/LagniezM17}.

Despite this significant progress, current model counters typically treat each problem instance in isolation, necessitating recomputation from scratch even when the constraints are syntactically similar and only minor modifications are made to the constraint set. 
This approach can be highly inefficient in iterative design or optimization processes where models are refined incrementally. 
For example, an infrastructure planner optimizing a road network may adjust only a handful of roads at a time, but must still recompute the model count for the entire network after each change. 

In this work, we present a model counting framework specifically engineered to exploit sequences of syntactically related problem instances. 
Specifically, we target a dynamic environment where the propositional theory evolves via arbitrary updates, including the addition or removal of clauses and modifications to the variable set. 
In this online setting, instances are processed incrementally, without the need for a lookahead, leveraging structural similarity to optimize performance.
Crucially, our framework requires weaker assumptions than Incremental SAT~\cite{DBLP:conf/sat/NadelR12}, accommodating broader structural modifications.
We elaborate on these distinctions and the implications for solver efficiency in Section~\ref{sec:rel_work}.

The core contribution of our work is a novel method for dynamically sharing computation knowledge across benchmarks. 
We focus on two forms of sharing: \emph{cache sharing} and \emph{branching heuristic sharing}. 
Achieving shared caching in this setting is non-trivial, which likely explains why, despite its clear potential, such an approach has not been realized previously. 
Classical caching techniques for model counting typically assume a static benchmark, where the underlying problem instance remains fixed. 
This assumption breaks down in our dynamic scenario, where formulas evolve across invocations. 
In particular, we show that effective cache reuse across different calls requires representing the entire formula in memory. 
Although this risks high memory consumption, our experiments show that cache reuse can yield orders-of-magnitude performance improvements over isolated strategies.
Further, we
 introduce a \emph{cache management strategy} utilizing quasi-canonical forms for symmetry detection. 
This significantly improves efficiency by identifying syntactically equivalent subproblems, as we demonstrate on selected benchmarks.
Additionally, we explore the reuse of branching heuristics, which affects the choice of which variable to branch on, during the model counting process. We show that sharing the same tree decomposition, that encapsulates variable ordering constraints, is especially effective on hard cases, in which subsequent formulas are obtained mostly by clause removal.

Our framework is implemented on top of \emph{d4}, a state-of-the-art model counter, featuring advanced cache management techniques suitable for our purposes. 
We focus our evaluation on the problems of \emph{dynamic argumentation}~\cite{DBLP:conf/comma/RotsteinMGS10} and \emph{soft cores computation}~\cite{DBLP:conf/icaart/AudemardLMR22} where syntactically similar formulas naturally arise and efficient handling is essential.
We evaluate our tool on a suite of synthetic and real-world benchmarks, originating from~\cite{argCompet,DBLP:journals/jea/FichteHH21,DBLP:conf/kr/LagniezL24}. Our results show that our tool achieves performance improvements over current leading tools, highlighting its effectiveness on syntactically similar benchmarks. 
%
%

Section~\ref{sec:rel_work} discusses related work, followed by preliminaries in Section~\ref{sec:prelims}. 
Section~\ref{sec:framework} formalizes the incremental framework. 
We present persistent caching in Section~\ref{sec:global_caching} and heuristic reuse in Section~\ref{sec:heur}. 
Experiments are detailed in Section~\ref{sec:evaluation}, followed by conclusions in Section~\ref{sec:conclusion}.

\section{Related Work}~\label{sec:rel_work}

We first distinguish our work from Incremental {SAT}.
In the SAT domain, incrementality typically refers to scenarios where the formula remains static but some literals are fixed as assumptions. 
In practice, allowing certain literals to be set as assumptions often suffices for applications that require the addition or removal of clauses~\cite{DBLP:conf/fmcad/Nadel10}. 
This is achieved by adding an auxiliary literal to each clause, which is set to false to activate the clause and true to deactivate it. 
Through this mechanism, SAT solvers can retain learnt clauses and internal data structures across different queries, as the base formula itself does not change: only the set of assumed literals does. 
Consequently, clause-learning and variable activity information can be preserved and reused, thereby improving performance on closely related instances, as is standard in incremental SAT solving.

A similar principle applies to model counting: recent work~\cite{DBLP:conf/aaai/LagniezM19} has shown that maintaining a persistent cache across multiple counting queries can offer significant performance benefits. 
This allows retaining learned clauses, caches, and variable weights, mirroring incremental SAT strategies to enable efficient information reuse across subproblems.
However, this form of incrementality is fundamentally limited. 
It is only practical when the set of all possible formulas is known in advance and when auxiliary variables can be introduced to activate or deactivate parts of the formula. 
If future modifications to the formula cannot be predicted, it becomes infeasible to preserve and reuse information between different model counting calls.

A recent model counter for Pseudo-Boolean (PB) formulas that, like our work, targets the incremental setting is
\texttt{PBCount2}~\cite{yang2024projectedincrementalpseudobooleanmodel}. 
There are, however, fundamental architectural distinctions between the two approaches.
Most notably, it relies on compiling constraints into Algebraic Decision Diagrams (ADDs). Its incremental mechanism caches these individual constraint ADDs, which must be retrieved via a linear scan and re-merged to reflect updates in the formula.
In contrast, our framework operates directly on the CNF structure using component caching. Rather than storing constraint fragments, we cache subformulas (components) alongside their computed model counts. This allows for immediate retrieval via hash-based lookups during search, bypassing the need for costly diagram reconstruction or merge operations.
Furthermore, while \texttt{PBCount2} performs a polynomial-time retrieval step once per formula update, our approach employs a constant-time hashing check at every node in the search tree, offering a different trade-off geared towards deep search performance.
Consistent with our methodology, \texttt{PBCount2} also disables preprocessing in incremental mode to preserve cache validity. We nevertheless include it in our discussion as it represents the primary existing framework for exact incremental counting, although its focus on PB constraints and ADDs differs from our CDCL-based CNF approach.
A major difference in the experimental settings between our work and \texttt{PBCount2} is that they have a 3-step or 5-step incremental model counting presented in their evaluation section. Namely, each benchmark undergoes 3 or 5 incremental changes. On the contrary, our argumentation setting deals with 1000 steps, while our soft core setting has $m$ steps, $m$ being the number of clauses in the input formula. Therefore, in our setting, although for the first 5 steps, \texttt{PBCount2} may outperform our tool, we handle the remaining many more steps, in which PBCount2 struggles.

Another related work worth mentioning in regards to our work is \texttt{Cara}~\cite{Illner_2025} which uses similar symmetry techniques in caching but for the purpose of knowledge compilation, rather than model counting. We further discuss the differences in the full version of our paper.

Despite the sophistication of modern counters, the majority remain strictly static, discarding all learned state between invocations. Aside from \texttt{PBCount2}, no existing tool supports the reuse of learned subproblems across sequences of syntactically similar CNF formulas. This gap motivates our contribution: a dedicated incremental CNF counter that enables effective cross-instance cache sharing without the overhead of full knowledge compilation.

\section{Background}\label{sec:prelims}

Let $\mathcal{L}$ be a propositional language constructed from a finite set of propositional variables $\mathcal{P}$ and the standard logical connectives.  
A \emph{literal} $\ell$ is either a propositional variable (e.g., $x$) or its negation ($\neg x$).  
Given a literal $\ell$ over a variable $x$, the \emph{complementary literal} $\overline{\ell}$ is defined as  $\overline{\ell} = \neg x$ if $\ell = x$, and $\overline{\ell} = x$ if $\ell = \neg x$. We denote the variable associated with a literal $\ell$ by $\mathit{Var}(\ell) = x$.  
A \emph{term} is a conjunction of literals, and a \emph{clause} is a disjunction of literals.  
Terms and clauses may, when convenient, be regarded as sets of literals.  
A CNF  formula $\form$ is a conjunction of clauses and may also be treated as a set of clauses when appropriate.  
$\mathit{Var}(\form)$ denotes the set of variables in $\form$.

\begin{example}~\label{ex:running}
Consider the CNF formula $\form$ over variables $\mathit{Var}(\form) = \{x_1, \ldots, x_5\}$ with clauses: $\{x_1 \vee x_2 \vee x_3, \neg x_1 \vee \neg x_2 \vee \neg x_3, x_4 \vee \neg x_1, x_5 \vee x_1, x_5 \vee x_2 \vee x_3, x_4 \vee \neg x_2 \vee \neg x_3\}$.
\end{example}

An \emph{interpretation} (or \emph{world}) over $\mathcal{P}$, denoted by $\omega$, is a mapping from $\mathcal{P}$ to $\{0, 1\}$.  
Interpretations are often represented as sets of literals, one per variable in $\mathcal{P}$, consisting of exactly those literals assigned the value $1$ by $\omega$.  
The set of all possible interpretations is denoted by $\mathcal{W}$.  
An interpretation $\omega$ is said to be a \emph{model} of a formula $\form \in \mathcal{L}$ if it satisfies the formula under the standard truth-functional semantics.  
The \emph{SAT problem} asks whether such a model exists.  
The set of all models of a formula $\form$ is denoted by $\text{mod}(\form)$ and is defined as  $\text{mod}(\form) = \{\omega \in \mathcal{W} \mid \omega \models \form\}$.  
The symbol $\models$ denotes logical entailment, and $\equiv$ denotes logical equivalence.  
For any formulas $\form, \Psi \in \mathcal{L}$, we have: $\form \models \Psi$ if and only if $\text{mod}(\form) \subseteq \text{mod}(\Psi)$ and $\form \equiv \Psi$ if and only if $\text{mod}(\form) = \text{mod}(\Psi)$. 
The notation $\norm{\form}$ denotes the number of models of $\form$ over the variables in $\mathit{Var}(\form)$.
For example, the number of models of $\form$, as defined in Example~\ref{ex:running}, is  $\norm{\form} = 10$. 
The problem of \emph{model counting} determines the number of models of a Boolean formula $\form$.

Since model counting is closely related to the SAT problem, many of the most effective strategies are built upon DPLL-based SAT solvers~\cite{DBLP:series/faia/GomesSS21}. 
These solvers are extended beyond satisfiability checking to compute the exact number of satisfying assignments. 
In particular, DPLL-based model counters recursively explore the search space by branching on variables and simplifying the formula at each decision point. 
Unlike SAT solvers that halt upon finding one solution, model counters must exhaustively enumerate all models, making the task significantly more challenging.

A standard optimization in DPLL-based counting is connected component decomposition. 
When the formula splits into variable-disjoint subformulas, their counts are computed separately and multiplied, drastically pruning the search space~\cite{DBLP:conf/sat/SangBBKP04}.
Another crucial optimization is component caching: because identical subproblems can arise in different branches of the search tree, model counters store the counts of previously solved components in a cache~\cite{DBLP:conf/sat/SangBBKP04,DBLP:conf/sat/Thurley06,cacheD4}. 
If the same component is encountered again, the solver can retrieve the cached result instead of recomputing it. 
Component decomposition and caching underpin state-of-the-art exact counters, enabling scalability on structured instances~\cite{DBLP:conf/sat/SangBBKP04,DBLP:conf/sat/Thurley06,DBLP:conf/cp/KorhonenJ21,DBLP:conf/ijcai/SharmaRSM19,DBLP:conf/ijcai/LagniezM17}.

In the heart of these optimizations, branching variable selection remains critical for performance. 
By exploiting structural properties, such as those derived from tree decompositions, heuristics can be guided to significantly reduce the effective search space~\cite{DBLP:conf/cp/KorhonenJ21}.

Nevertheless, all the mechanisms employed by a model counter (such as component decomposition, caching, and branching heuristics) are intrinsically tied to the specific instance being solved. As a result, these structures and strategies cannot typically be reused directly when addressing a new problem instance. This paper challenges this situation. As we demonstrate in the following sections, it is possible, under certain conditions, to fully reuse the cache structure regardless of the subsequent formula. 

\section{The Incremental Model Counting Framework~\label{sec:framework}}

In this work, we formalize the problem of incremental model counting as a dynamic process operating on an evolving propositional theory. 

Unlike the static setting, our framework maintains a current state that is modified by sequences of updates. 

Crucially, model counting is not performed after every atomic modification, but rather at specific checkpoints after a batch of operations has been applied.

The framework we describe in the following subsection is a general framework that is intended to be flexible to capture the fact that the definition of “incremental” is somehow from the user’s perspective. In some benchmarks (e.g. soft core),  “incremental” can mean changing a single clause, while in others (e.g. argumentation), "incremental” means much wider changes. We do not want to limit our framework to allow for this flexibility. 

\subsection{State and Operations}

Let $\mathcal{U}$ be the universe of all propositional variables. We define the 
state of the system at time step $t$ as a pair $\Phi_t = \langle \mathcal{V}_t, \mathcal{C}_t \rangle$, 
where $\mathcal{V}_t \subseteq \mathcal{U}$ is the current set of active variables 
and $\mathcal{C}_t$ is a set of clauses over $\mathcal{V}_t$. The model count of 
the state $\Phi_t$ is defined as the number of satisfying assignments of the CNF 
formula formed by $\mathcal{C}_t$ over the variables $\mathcal{V}_t$.

The transition between states, denoted $\Phi_{t+1} = \text{update}(\Phi_t, \delta)$, 
is driven by an update operation $\delta$. We consider the following four atomic rules for the update operation:

\begin{itemize}
    \item \textit{Clause Addition} ($\text{add\_clause}(c)$): Adds a new clause $c$ to the theory: 
    $\mathcal{V}_{t+1} = \mathcal{V}_t, \mathcal{C}_{t+1} = \mathcal{C}_t \cup \{c\}$.
    \textit{Precondition:} $vars(c) \subseteq \mathcal{V}_t$.
    
    \item \textit{Clause Removal} ($\text{rem\_clause}(c)$): Removes an existing clause from the theory:
    $\mathcal{V}_{t+1} = \mathcal{V}_t,\mathcal{C}_{t+1} = \mathcal{C}_t \setminus \{c\}$.
    
    \item \textit{Variable Addition} ($\text{add\_var}(v)$): Introduces a new variable $v$ into the scope:
    $\mathcal{V}_{t+1} = \mathcal{V}_t \cup \{v\}, \mathcal{C}_{t+1} = \mathcal{C}_t$.
    \textit{Note:} Adding a variable without adding constraints essentially doubles the model count.
    
    \item \textit{Variable Removal} ($\text{rem\_var}(v)$): Removes a variable $v$ from the scope:
    $\mathcal{V}_{t+1} = \mathcal{V}_t \setminus \{v\}, \quad \mathcal{C}_{t+1} = \mathcal{C}_t$.
    \textit{Precondition:} The variable $v$ must not appear in any clause in $\mathcal{C}_t$. 
    Formally, $\forall c \in \mathcal{C}_t, v \notin vars(c)$. This ensures that we do not 
    leave ``dangling'' literals in the clause database.
\end{itemize}

\paragraph{Problem Definition}
\label{sec:problem_formulation}
We define an \textit{update batch}, denoted as $\Delta$, as a finite sequence of atomic operations.
%

The \textit{Incremental Model Counting} problem is defined as follows: Given an initial state $\Phi_0 = \langle \mathcal{V}_{init}, \mathcal{C}_{init} \rangle$ 
and a sequence of update batches $\boldsymbol{\Delta} = \langle \Delta_1, \Delta_2, \dots, \Delta_k \rangle$, 
compute the sequence of counts: \[ \langle \norm{\Phi_1}, \norm{\Phi_2}, \dots, \norm{\Phi_k} \rangle \]
where $\Phi_i$ is the state resulting from applying the sequence of operations in $\Delta_i$ to $\Phi_{i-1}$.


%
The objective is to minimize the total computational time of the entire sequence of counts. 
We achieve this by leveraging information learned during the computation of $\Phi_{0 \dots i-1}$ to accelerate the counting of $\Phi_i$.

\section{Shared Caching in Model Counting}\label{sec:global_caching}

As detailed in the preliminaries, model counters employ search procedures that recursively decompose the input formula~$\Sigma$ into subformulas. 
To optimize this process, computed counts are stored in a cache: if a subformula~$\Sigma'$ is encountered again during the search, its model count is retrieved directly from the cache, thereby eliminating the need for redundant re-computation.

Our strategy persists the cache across benchmarks to enable subformula count reuse.
While conceptually simple, this presents significant implementation challenges. 
Section~\ref{sec:cachModCounters} analyzes the limitations of standard caching mechanisms in this context, while Section~\ref{sec:cacheLazy} introduces a novel optimization to adapt the basic scheme for efficient reuse.

\subsection{Cache in Model Counters}~\label{sec:cachModCounters}

During the search, the solver encounters subformulas $\varphi$ derived from $\Sigma$ via conditioning. 
We define a \emph{caching scheme} $c$ as a mapping from each $\varphi$ to a representation $r_c(\varphi)$, with the cache storing the map $r_c(\varphi) \mapsto \norm{\varphi}$. 
A cache miss occurs if no matching representation is found. 
A scheme is \emph{correct} if collision implies logical equivalence: $r_c(\varphi_1) = r_c(\varphi_2) \implies \varphi_1 \equiv \varphi_2$.

The performance of a caching scheme is determined by a trade-off between memory footprint and the granularity of equivalence detection. 
We review three standard approaches and analyze their applicability to our dynamic setting.
In the basic scheme used by {\tt cachet}, each subformula $\varphi$ is represented explicitly as a string of literals, with clauses separated by zeroes. While robust, this representation is memory-intensive.
To improve efficiency, {\tt SharpSAT} employs a hybrid scheme. 
It assigns static indices to the clauses of the input formula $\Sigma$. 
A subformula is represented by a pair: the set of variables present and the sorted list of indices of the \emph{original} clauses that belong to the subformula. 
Crucially, to save space, {\tt SharpSAT} omits indices of binary clauses from this representation, relying on the static implication graph to handle them implicitly.
{\tt d4} adopts a structured explicit approach. 
It stores the clauses of $\varphi$ directly but applies aggressive normalization: literals within clauses and the clauses themselves are sorted lexicographically to maximize cache hits. 
To optimize memory, {\tt d4} does not store clauses that were originally binary in $\Sigma$, nor clauses that have not been shortened by the current assignment.

The optimized schemes of {\tt SharpSAT} and {\tt d4} fundamentally rely on the assumption that the input formula $\Sigma$ is static. They achieve compression by treating binary clauses (or unshortened clauses) as implicit background knowledge that does not need to be part of the cache key.

In our incremental setting, however, where the formula evolves, this assumption leads to unsoundness.
Consider two different formulas: $\Sigma_1 = \{x_1 \vee x_2\}$ and $\Sigma_2 = \{x_1 \vee x_2, \neg x_1 \vee \neg x_2\}$.
Under the schemes of {\tt SharpSAT} or {\tt d4}, the cache representation of both $\Sigma_1$ and $\Sigma_2$ would be identical: a variable set $\{x_1, x_2\}$ with an empty clause list (since the clauses are binary and thus omitted).
However, $\Sigma_1$ has 3 models while $\Sigma_2$ has 2. A cache entry created for $\Sigma_1$ would be incorrectly retrieved for $\Sigma_2$, yielding a wrong count.

Consequently, we cannot use these compressed representations. We must adopt a fully explicit caching scheme similar to {\tt cachet}, which captures the exact semantic content of the subformula regardless of the global context. In the following section, we introduce optimizations to mitigate the memory overhead of this approach.

\subsection{The Lazy Symmetry Caching}\label{sec:cacheLazy}

As discussed in the previous section, we decided  to employ an explicit caching scheme to ensure correctness. 
To maximize efficiency, we adopt {\tt d4}'s normalization: literals and clauses are lexicographically sorted, and duplicates removed.
Crucially, our method goes beyond exact variable matching, detecting equivalent subproblems even when defined on disjoint variable sets. 
In the following, we introduce \emph{symmetry caching} and detail our novel \emph{lazy symmetry caching} mechanism, which exploits our explicit representation to identify these symmetries.

Symmetries are typically exploited via permutations: a permutation $\sigma$ over $L_{\mathcal{P}}$, the set of all literals over $\mathcal{P}$, is a bijective mapping from $L_{\mathcal{P}} = \mathcal{P} \cup \{\neg x \mid x \in \mathcal{P}\}$ to itself.
Any such permutation $\sigma$ can be naturally extended to a homomorphism on propositional formulas, such that for every Boolean connective $c$ of arity $k$: 
$\sigma(c(\alpha_1, \ldots, \alpha_k)) = c(\sigma(\alpha_1), \ldots, \sigma(\alpha_k))$.

Throughout this paper, we assume that any permutation $\sigma$ under consideration satisfies the following \emph{stability condition}: for any pair of literals $\ell_1, \ell_2$, it holds that $\sigma(\ell_1) = \ell_2$ if and only if $\sigma(\overline{\ell_1}) = \overline{\ell_2}$.

Each permutation $\sigma$ is represented in \emph{simplified cycle notation}: as a product of cycles corresponding to its nontrivial orbits (i.e., those of length at least two). In this notation, for each orbit $\{ \ell_1, \ldots, \ell_k \}$, only one of the two cycles $(\ell_1 \ldots \ell_k)$ or $(\overline{\ell_1}~\ldots~\overline{\ell_k})$ is shown. For example, if $L_{\mathcal{P}} = \{x_1, \ldots, x_6\}$, the permutation represented by $(x_1~\neg x_3~x_4)(x_5~x_6)$ corresponds to mapping $x_1$ to $\neg x_3$, $\neg x_1$ to $x_3$, $x_3$ to $\neg x_4$, $\neg x_3$ to $x_4$, $x_4$ to $x_1$, $\neg x_4$ to $\neg x_1$, $x_5$ to $x_6$, $\neg x_5$ to $\neg x_6$, $x_6$ to $x_5$, and $\neg x_6$ to $\neg x_5$, with $x_2$ and $\neg x_2$ remaining fixed. The identity permutation is represented by the empty product in this notation.

\begin{example}[Example~\ref{ex:running} cont'd]\label{ex:sym}
Consider the formula $\form$ from Example~\ref{ex:running}. Suppose we evaluate the assignment $\{x_3\}$, resulting in the residual formula
$\varphi_{x_3} = \{\neg x_1 \vee \neg x_2,\ x_4 \vee \neg x_1,\ x_5 \vee x_1,\ x_4 \vee \neg x_2\}$.
Now, consider instead the assignment $\{\neg x_3\}$, for which the residual formula is
$\varphi_{\neg x_3} = \{x_1 \vee x_2,\ x_4 \vee \neg x_1,\ x_5 \vee x_1,\ x_5 \vee x_2\}$.
Let $\sigma = (x_1~\neg x_1)(x_2~\neg x_2)(x_4~x_5)$ be the permutation that swaps $x_1$ with $\neg x_1$, $x_2$ with $\neg x_2$, and $x_4$ with $x_5$. Note that applying $\sigma$ to $\form$ yields $\sigma(\varphi_{x_3}) = \varphi_{\neg x_3}$.
\end{example}

As shown in~\cite{DBLP:conf/ecai/BartKLM14}, in the context of knowledge compilation, symmetry caching is a powerful technique that can significantly reduce the size of the representation.
The authors propose a greedy method that searches the cache for an entry $\varphi_s$ equivalent to the current formula $\varphi$ up to variable renaming, filtering candidates by size and literal counts. 
However, this approach is computationally expensive due to the cost of verifying renamings, and relies on generic hash properties that increase collision rates.

To overcome the limitations of symmetry caching over shared caching, we propose a lazy approach in which each formula inserted into the cache is stored by using a canonical representation. 
Rather than the standard lexicographic ordering, we employ a representation that incorporates structural information about the formula. 
We describe our method in steps. First, for each literal, we count the number of clauses in which it appears, and for each pair of literals, we sort them by their frequency of occurrence (increasing order). In the case of a tie, the negative literal is listed first.

\begin{example}[Example~\ref{ex:sym} cont'd]\label{ex:sortLit}
Consider the two formulas computed in Example~\ref{ex:sym}.  
For $\varphi_{x_3}$ , the literal pairs (for each variable) are: $(x_1, \neg x_1)$, $(x_2,\neg x_2)$, $(\neg x_4, x_4)$, and $(\neg x_5, x_5)$.  
For $\varphi_{\neg x_3}$ , the corresponding pairs are: $(\neg x_1, x_1)$, $(\neg x_2, x_2)$, $(\neg x_4, x_4)$, and $(\neg x_5, x_5)$.
\end{example}

We then sort these pairs (each associated with a variable) lexicographically according to the frequency of occurrence of each literal in the formula, with ties broken by lexicographical order.  For example note that the encoding of $(\neg x_4, x_4)$ in $\varphi_{x_3}$ is $(0,2)$.

\begin{example}[Example~\ref{ex:sortLit} cont'd]\label{ex:varOrder}
For $\varphi_{ x_3}$, after sorting, we obtain the ordering:
$(\neg x_5, x_5)$, $(x_2, \neg x_2)$, $(\neg x_4, x_4)$, $(x_1, \neg x_1)$. 
For $\varphi_{\neg x_3}$ , the sorted order is: $(\neg x_4, x_4)$, $(\neg x_2, x_2)$, $(\neg x_5, x_5)$, $(\neg x_1, x_1)$.

\end{example}

Finally, for each pair, we select the first literal and use its position in the sorted ordering to construct the associated permutation~$\sigma$. 
Specifically, for each variable, we consider the negative literal when associating positions in the permutation. 
In this way we normalize the formulas to have the same canonical ordering. 
For example in $\varphi_{\neg x_3}$, $(\neg x_4,x_4)$ is the \emph{first} pair in the order so the permutation $\sigma_{\neg x_3}$ takes $\neg x_4$ to $\neg x_1$, which is the \emph{first} (negated) literal. Similarly the permutation $\sigma_{x_3}$ takes $\neg x_5$ to $\neg x_1$.

\begin{example}[Example~\ref{ex:varOrder} cont'd]\label{ex:sigma}
Continuing with the formulas from Example~\ref{ex:varOrder}:  
For $\varphi_{x_3}$, we obtain $\sigma_{x_3} = (\neg x_4~\neg x_1)(\neg x_2~\neg x_2)(\neg x_5~\neg x_3)(\neg x_1~\neg x_4)$.  
For $\varphi_{\neg x_3}$, we obtain $\sigma_{\neg x_3} = (\neg x_5~\neg x_1)(x_2~\neg x_2)(\neg x_4~\neg x_3)(x_1~\neg x_4)$.
\end{example}

Once the permutation associated with a formula is determined, we store the formula obtained by applying the permutation in the cache, rather than the original formula. Since applying a permutation to a formula preserves the number of satisfying assignments, we all in all get that the permuted formula is associated with the model count of the original, non-permuted formula.

\begin{example}[Example~\ref{ex:sigma} cont'd]
Consider the two formulas from Example~\ref{ex:sym} along with their permutations computed in Example~\ref{ex:sigma}.  
We have $\sigma_{\neg x_3}(\varphi_{ x_3}) = \{x_4 \vee x_2,\ x_1 \vee \neg x_4,\ x_3 \vee x_4,\ x_3 \vee x_2\} = \sigma_{ x_3}(\varphi_{\neg x_3})$.
\end{example}

As this example illustrates, lazy symmetry caching allows us to identify and reuse formulas with identical model counts, even if they are not syntactically equivalent. In this approach, the hash function operates on the permuted (canonicalized) version of the formula. Since we already store the full formula in the cache, this strategy incurs no additional storage cost, aside from the modest computational overhead of determining the appropriate permutation.

It is important to note that this caching strategy is only directly applicable to model counting, where permutations do not affect the count. In the context of weighted model counting, additional care must be taken to ensure that the weights are also preserved under the applied permutation.

\section{Shared Heuristics }~\label{sec:heur}

The decision of which variable to branch on next is a critical factor that significantly affects the practical performance of model counters.
Numerous classical SAT branching heuristics have been proposed in the literature, such as MOM~\cite{DBLP:conf/dimacs/DuboisABC93}, JWTS~\cite{DBLP:journals/amai/JeroslowW90}, and VSIDS~\cite{DBLP:conf/dac/MoskewiczMZZM01}, along with heuristics specifically designed for model counting, including DLCS and VSADS~\cite{DBLP:conf/sat/SangBK05}. 
However, the most effective heuristics often exploit structural properties of the given formula.
In particular, the structure of a formula, captured by its primal graph, offers valuable information that can be leveraged to design more efficient decision heuristics.
By utilizing tree decompositions (TDs) of the primal graph, a solver can exploit this inherent structure, thereby potentially reducing the effective search space and substantially improving the efficiency of model counting.

A \emph{primal graph} of a CNF formula $\form$ is an undirected graph whose vertices are the variables of~$\form$, with edges between variables that appear together in at least one clause. 
A \emph{tree decomposition}~\cite{DBLP:journals/jal/RobertsonS86,DBLP:conf/sofsem/Bodlaender05} of a graph $G$ is a tree $T$ whose nodes, called \emph{bags}, are subsets of vertices of $G$, and which collectively cover the vertices and edges of $G$ while guaranteeing a certain connectivity property (see, e.g.,~\cite{DBLP:journals/jal/RobertsonS86}). The \emph{width} of a decomposition is the size of its largest bag minus one; the \emph{treewidth} of $G$ is the minimum such width. 

Tree decomposition is fundamental in model counting, enabling dynamic programming algorithms with time complexity $\mathcal{O}(2^k \cdot n)$, where $k$ is the treewidth. 
This makes the problem fixed-parameter tractable w.r.t. $k$, offering efficiency for low-treewidth formulas. 
However, since practical instances often have large treewidths, state-of-the-art counters use this structural information more flexibly. 
For instance, Korhonen and Järvisalo~\cite{DBLP:conf/cp/KorhonenJ21} show that \emph{hybrid scores} (combining tree decomposition data with traditional heuristics) improve performance even when theoretical bounds are not strictly met.

This hybrid approach enables modern model counters to exploit the advantages of structural decomposition even for instances of high treewidth, by strategically prioritizing variables according to their structural relevance.
In our setting, we consider the model counting problem over a repository of multiple, closely related Boolean formulas. 
Since computing a tree decomposition can be computationally expensive, performing this preprocessing step separately for each formula in the repository is generally impractical. 
Instead, we propose computing the tree decomposition only once, during the initial iteration, and reusing it for all subsequent formulas in the repository. 
This approach amortizes the overhead of tree decomposition construction, reducing it from $N$ computations (one per formula) to a single computation per repository, and thus yields significant efficiency gains by eliminating redundant work and enabling more consistent variable ordering across related instances. 

The key insight is that when a formula is modified solely by clause removal, its existing tree decomposition remains a valid structural representation, even if it is no longer optimal.
Moreover, as tree decompositions encode variable ordering constraints, integrating this structural information into the branching heuristic guides the solver to explore similar subformulas when analyzing related formulas. Empirical results, presented in the next section, confirm that this strategy can yield significant performance improvements.

For similar reasons, we opt for the DLCS heuristic rather than VSADS, the latter being generally more effective when solving a single, standalone benchmark. 
The DLCS heuristic assigns each variable a weight equal to the number of clauses in which it appears and selects the variable with the highest score for branching. 
In contrast, VSADS multiplies the DLCS score of a variable~$v$ by the number of times $v$ participates in conflicts encountered during the search, thus incorporating conflict-driven information. 
Our intuition is that DLCS, being less influenced by conflict-driven clause learning, yields more stable and predictable branching decisions across syntactically similar formulas, thereby increasing the likelihood of positive cache hits. 
Experimental results support this hypothesis, showing that DLCS consistently outperforms VSADS in our setting.
\section{Implementation and Evaluation}\label{sec:evaluation}

In this section, we present two evaluation settings. The first examines the benefits of sharing when counting the number of complete extensions in a dynamic argumentation framework~\cite{DBLP:conf/ijcai/Dung93}. The second setting addresses the computation of subset-minimal soft cores~\cite{DBLP:conf/icaart/AudemardLMR22}. We describe each setting separately, explain how it fits into our incremental model counting framework, and show by evaluation how the re-use of knowledge across the various instances succeeds over methods in which each instance is a standalone.

%


\subsection{Settings and Research Questions}

All the experiments for both settings were carried out on a cluster equipped with dual 32-core Intel\textsuperscript{\textregistered} Xeon\textsuperscript{\textregistered} Gold 6130 CPUs (2.10GHz) and 32~GiB of memory, running Rocky Linux 8.7 (kernel version 4.18.0-425.13.1.el8\_7.x86\_64) and using the \texttt{gcc} 11.4.0 compiler. All code was implemented in C++. We enforce a timeout limit of 60 minutes per benchmark.

For our experiments, we developed a tool named \ourtool, based on the state-of-the-art model counter $\mathtt{d4}$\footnote{\url{https://github.com/crillab/d4v2}}. 
\ourtool\ implements the incremental framework formalized in Section~\ref{sec:framework}, evolving the formula through step-by-step modifications while maintaining a persistent state, rather than solving a sequence of independent instances.
Our tool supports three caching modes:
\emph{No-Sharing ({\tt NoShared})}, where each instance is solved independently with a local cache;
\emph{Shared-cache ({\tt Shared})}, where the cache persists across updates; and
\emph{Shared cache + Symmetry ({\tt Shared+Sym})}, which extends the shared mode by activating \ourtool's symmetry detection during cache operations. See Section~\ref{sec:global_caching} for details.
Additionally, \ourtool\ features an orthogonal \texttt{Tree Decomposition Sharing (TD)} mode, as explained in Section~\ref{sec:heur}, which reuses the tree decomposition computed for the initial formula for all subsequent instances. Finally, the tool supports two branching heuristics, {\tt VSADS} and {\tt DLCS}, as detailed in Section~\ref{sec:heur}.

For each setting our objective is to evaluate the following. First (RQ1), how does the deterministic {\tt DLCS} branching heuristic perform compared to the {\tt VSADS} branching heuristic, across all of our tool modes. Second (RQ2), which mode of \ourtool with the best branching heuristic, is the most efficient.  Finding that either of the shared modes is the most efficient, demonstrates the advantage of knowledge re-use for our problem settings. Finally (RQ3), we are also interested to learn how this best mode is compared against other state-of-the-art model counters. For that, We benchmark against two external solvers: \ganak~\cite{DBLP:conf/cav/SoosM25}, the winner of the 2025 Model Counting Competition,\footnote{\url{https://mccompetition.org/}} and PBcount2~\cite{DBLP:conf/sat/Yang0M25}, which similarly to us, employs knowledge reuse across instances. These tools were executed using their default (or near-default) settings, following common practice.

\subsection{Argumentation}\label{sec:argumentation}

\subsubsection{Problem Definition}

To thoroughly evaluate our approach, we adopt as a test bed the problem of counting extensions within an Argumentation Framework (AF), focusing particularly on the complete semantic of dynamic argumentation, wherein the AF may evolve over time.

\emph{Abstract argumentation}, introduced by Dung~\cite{DBLP:conf/ijcai/Dung93}, provides a foundational paradigm for formalizing argumentative reasoning in artificial intelligence. An \emph{argumentation framework} consists of a set of abstract arguments together with binary attack relations among them. The principal computational task is to identify subsets of arguments, called \emph{extensions}, which are deemed collectively acceptable according to different \emph{semantics}.
Formally, a set of arguments $S$ constitutes a \emph{complete extension} if: (i) $S$ is conflict-free (no arguments in $S$ attack each other); (ii) every argument in $S$ is defended by $S$ against attacks from outside; and (iii) $S$ contains all arguments that it defends.
Beyond classical acceptance tasks, the \emph{counting} problem in abstract argumentation involves quantifying the number of extensions under a given semantics~\cite{iccma21}, offering insight into the diversity and nature of solutions. This is particularly valuable in scenarios such as preference-based decision making and probabilistic reasoning.

\emph{Dynamic argumentation} examines how argument status shifts under modification, a critical concern in evolving multi-agent systems~\cite{doi:10.3233/AAC-230013}.
To systematically evaluate performance, we employ \texttt{crustabri}~\cite{crustabri_github} to generate dynamic Argumentation Frameworks (AFs) by iteratively applying random perturbations: 
 \begin{itemize}
    \item \textit{Delete argument} (10\%): Remove an argument and its attacks.
    \item \textit{Add argument with attacks} (20\%): Add a new argument that attacks others.
    \item \textit{Delete attacks} (40\%): Remove random attack relations.
    \item \textit{Add attacks} (30\%): Add new attack relations between existing arguments.
\end{itemize} 

Each resulting AF is translated into a CNF formula $\Gamma$ using the standard encoding from~\cite{DBLP:conf/nmr/BesnardD04} and implemented in \texttt{crustabri}. We adapt our incremental framework terminology from Section~\ref{sec:framework}. However, given the potentially extensive structural differences between consecutive AFs, we implement the transition from $\form_{i-1}$ to $\form_i$ as a full reset: the update batch clears the current state and inserts all clauses of $\Gamma$.
Consequently, each initial AF benchmark yields a sequence $\langle \form_1, \dots, \form_{1000} \rangle$ of CNF instances derived from the evolving framework.

For a dataset, we use 264 AFs from the 2023 Abstract Argumentation Competition\cite{argCompet}. The AFs are described in Appendix~\ref{app:afFamilies}.
We have used the {\tt Cleaning Expectation} method of {\tt d4} to prevent memory-outs.





\subsubsection{Evaluation Details}\label{sec:res_Argumentation}


\subsubsection{Answering RQ1}

To evaluate the impact of the branching heuristic on the effectiveness of our approach, we performed a comparison between DLCS and VSADS on all benchmarks and modes.
%
%
%
%
%
\ourtool\ with {\tt DLCS} reaches a peak of 203 solved benchmarks, outperforming {\tt VSADS}, which tops out at 172. 
Specifically, {\tt DLCS} solves 187, 188, and 203 benchmarks across the {\tt No Shared}, {\tt Shared}, and {\tt Shared+Sym} configurations, respectively, while {\tt VSADS} solves 166, 172, and 171 in the same settings.

When analyzing the results, we observed that, on average, {\tt DLCS} is approximately 15\% faster than {\tt VSADS} in our experimental setup.
%
%
This outcome is consistent with our intuition: the deterministic nature of {\tt DLCS} increases the likelihood of positive cache hits.
Since caching relies on re-encountering similar substructures during the search, a deterministic traversal strategy such as {\tt DLCS} promotes more consistent exploration patterns, thereby enhancing cache reuse and improving overall efficiency of \ourtool. As such, we decided to continue with the DLCS heuristic for our remaining comparisons.




\subsubsection{Answering RQ2}

\begin{figure}[t!]
    \centering
    \includegraphics[width=0.9\columnwidth]{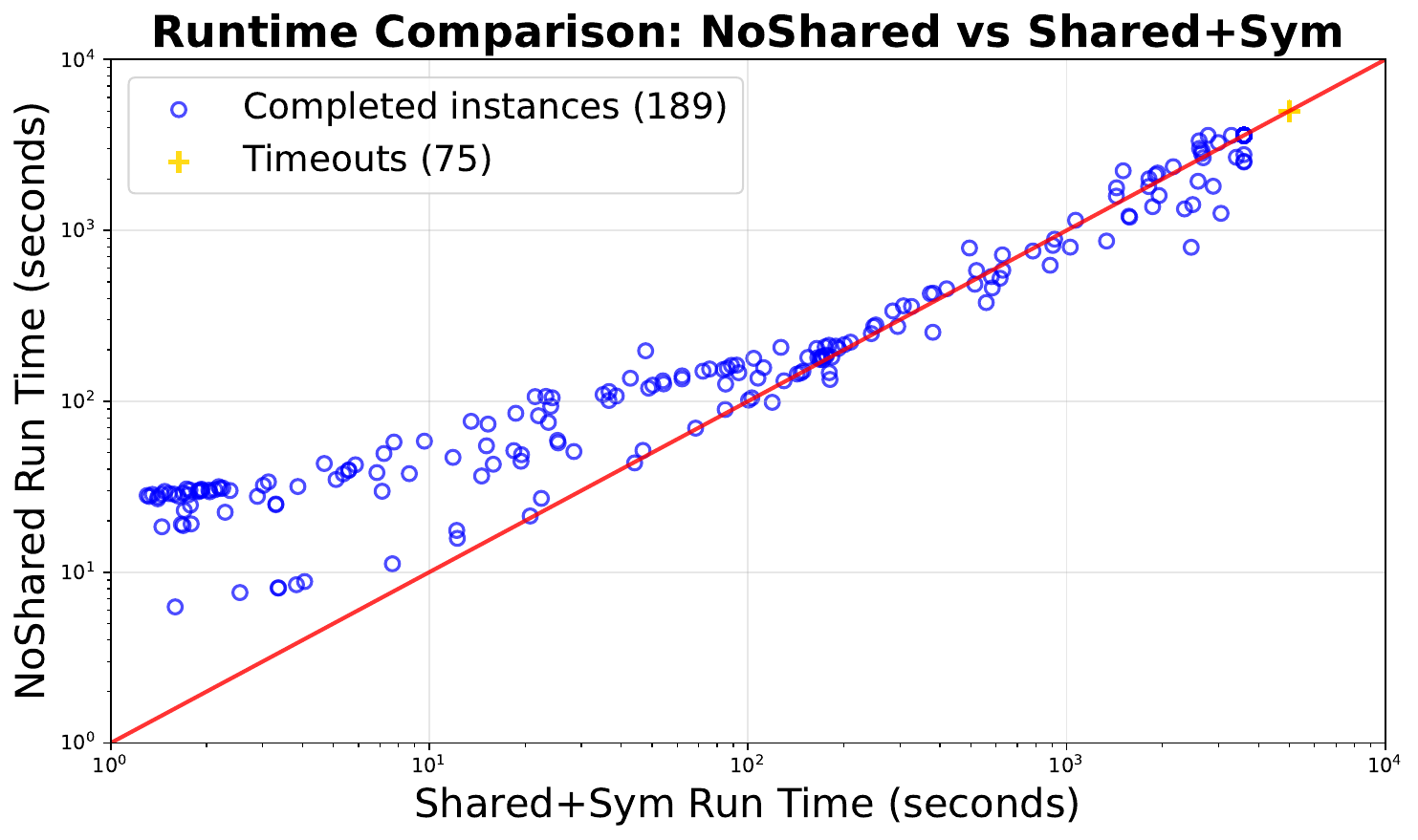}
    \caption{\label{fig:scatterArgNS-SS}Comparing the runtimes of \ourtool({\tt NoShared}, {\tt DLCS}) and \ourtool({\tt Shared+Sym}, {\tt DLCS})}
   \vspace{-15pt}
\end{figure}


We ran \ourtool with all three modes. Since the modifications between formulas are random, we observed that shared Tree Decomposition (TD), does not seem to scale, and indeed after preliminary tests we decided not to include TD in this comparison.
This is likely because the tree decomposition computed for $\form_i$ becomes structurally unsuitable for $\form_{i+1}$ due to the significant changes in the underlying argumentation framework.

Moreover, we observed that \texttt{Shared} vs. \texttt{Shared+Sym} produces almost the same results (see below for more on that). 
As such, we focus on the comparison between \texttt{NoShared} and \texttt{Shared+Sym} configurations across all benchmarks, as can be seen in Figure~\ref{fig:scatterArgNS-SS}. 
Each point in the plot corresponds to a sequence of formulas derived from a specific AF benchmarks.
The y-axis shows the runtime of \ourtool in the {\tt NoShared} mode, while the x-axis shows the runtime in the {\tt Shared+Sym} mode.

As can be seen, in most cases {\tt Shared+Sym} yields a substantial performance gains, with a total improvement rate of 5000\%  on average (1200\% median).
We notice that this gain is particular for faster repositories. Specifically, improvement could vary between 20000\% on faster benchmarks to almost even a minor slow down of a couple of percents on more complex or slower benchmarks.
We suspect that this behavior arises because the benefits of caching diminish as benchmarks increase in complexity. 
In particular, the memory cost of maintaining the full formula context in the shared caching scheme can be prohibitive compared to the {\tt NoShared} mode, which generally maintains a significantly smaller cache footprint.

For {\tt Shared} vs {\tt Shared+Sym}, the difference was minor and on average is 4\% in favor of using symmetry. We suspect that the small gain of using symmetry is that the 1000 instances in each repository are similar, with the same variable names (since all benchmarks originate from the same source). In these cases, using symmetry that  intends to exploit similar structures between benchmarks,  differ only in the variables names, is mostly redundant. 

\subsubsection{Cache-hits Analysis}

To better understand the impact of shared caching, we analyze a specific AF benchmark: 
$afinput\_exp\_cycles\_indvary2\_step1\_batch\_yyy01$.
This benchmark comprises a sequence of a 1000 formulas, each with a small model count (2, 3, or 4).
While the {\tt NoShared} mode times out after solving 921 benchmarks, both shared caching modes successfully complete the entire sequence.
Specifically, {\tt Shared} and {\tt Shared+Sym} achieve runtime improvements of 78\% and 121\%, respectively, over the {\tt NoShared} baseline.

Our observation is that in all three modes of operation, the first iteration takes roughly the same runtime. As the run progresses, however, time differences start to become evident in favor of the shared-cache modes. To understand the reason behind this behavior, we checked the total positive and negative cache hits (= cache hits / misses, resp.) for the entire run. 
Regarding positive hits,  there are $0$ positive cache hits during the entire run of the {\tt NoShared} mode on this repository, opposed to 766 on the Shared mode and 764 on {\tt Shared+Sym}.
On the other hand, for negative hits, there are 71691 negative hits on the {\tt NoShared} mode on this repository, opposed to only 19758 ({\tt Shared}) and 17588 ({\tt Shared+Sym}). 
This implies that the runtime of the shared modes should indeed become faster as less branching and recursion is done on the formulas in order to count their models. 

%



\subsubsection{Answering RQ3}
\begin{figure}[t!]
    \centering
        \includegraphics[width=\columnwidth]{
        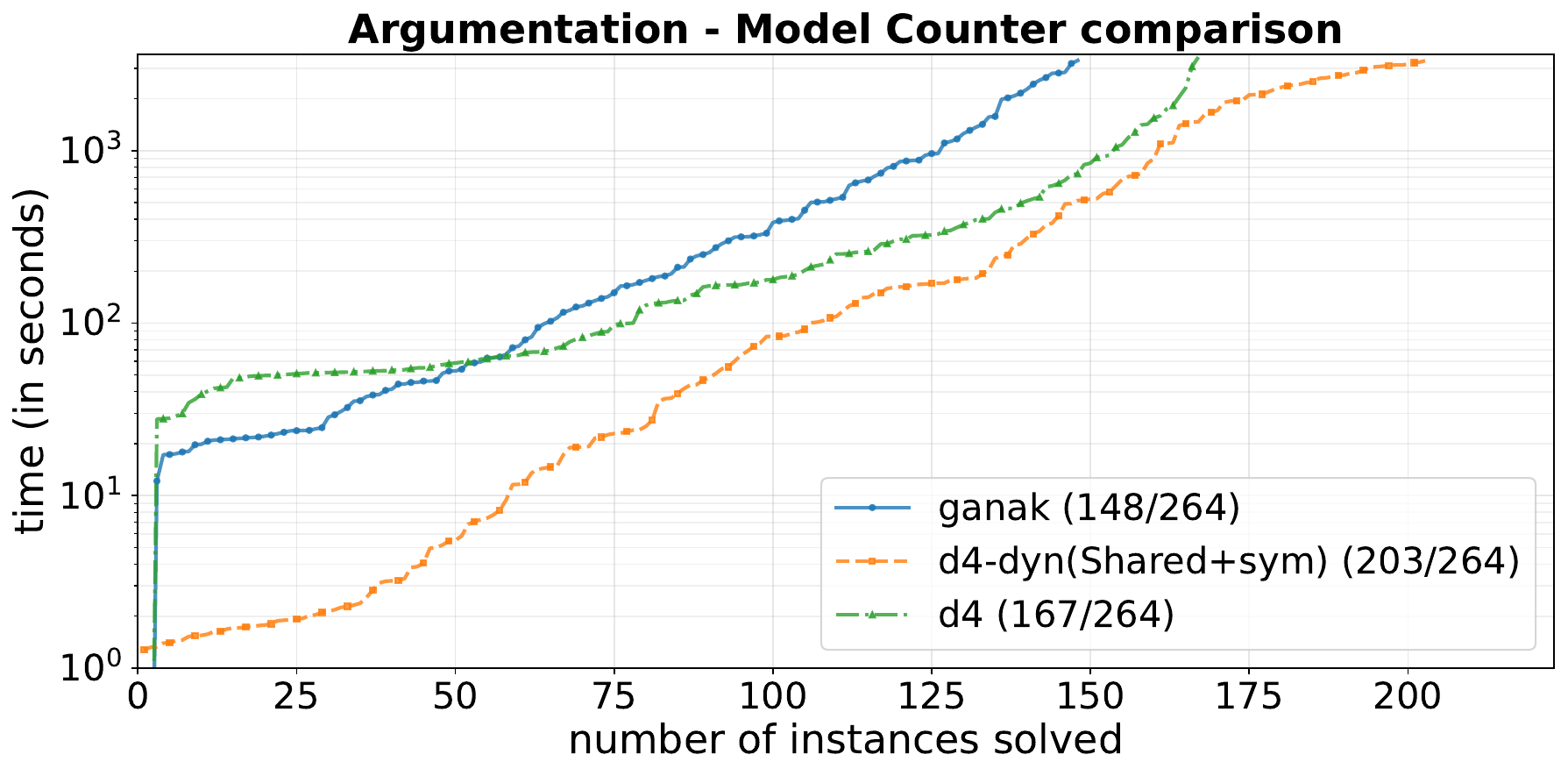}
        \caption{\label{fig:sub4}Evaluating the runtime of each method.}
        \vspace{-15pt}
    \end{figure}

We ran our tool against other state-of-the-art tools.
For that we executed baseline model counters on each repository of argumentation frameworks using a simple {\tt bash} script loop. 
Our evaluation comprises the following three experimental setups shown in Figure \ref{fig:sub4}.
We aimed to run the baseline tools with their default setting: (i) our tool \texttt{d4-dyn} (orange) on a Shared+Sym mode with DLCS branching heuristic; (ii) \texttt{d4}~\cite{DBLP:conf/ijcai/LagniezLM16} (green) with \emph{Glucose} as a SAT solver, flow cutter disabled and no preprocessing; (iii) \texttt{ganak} (blue) with default settings. 


    
       
    

We also tried comparing with {\tt PBCount2} but were limited to its Regular mode, as its Incremental mode is incompatible with our {\tt crustabri}-based batch workflow. After translating CNF files to the required .opb format, PBCount2  (Regular) solved 50 of 264 benchmark families, well below the other tools and therefore is not shown in the plot.

Overall, \ourtool\ solves 203 benchmarks, compared to 167 for {\tt d4} and 148 with \ganak. 
We must qualify these results by noting a structural advantage: \ourtool\ operates as a persistent process, incurring initialization costs only once. 
In contrast, competing solvers are re-invoked for each formula, accumulating significant overhead from repeated process instantiation.
%

\subsection{Soft Cores}
\label{sec:softcores}


\paragraph{Problem Definition}
\label{sec:softcore_def}
The notion of a soft core was introduced in~\cite{DBLP:conf/icaart/AudemardLMR22} as a concept to explain why some system behaviors are exhibited while others are excluded. A \emph{soft core} is a subset of constraints that is both small and highly constrained, meaning it admits only a limited number of models. Identifying a soft core is naturally a bi-criteria optimization problem, where both the size of the subset and the number of models are relevant. 
Depending on the context, one may prioritize minimizing soft core size or reducing model count. 
However, computing a minimal soft core is NP$^{\text{PP}}$-hard~\cite{DBLP:journals/jair/LittmanGM98}, making optimal solutions already infeasible for instances with more than 10 variables.

In our work, we consider a slightly modified variant, which searches for a minimal subset soft core. To achieve this, we employ a simple iterative strategy. For that recall that removal of clauses increases the model-count. Starting from the original formula, we do a single iteration over all clauses, in which we remove one clause at a time and check the new model count. If that model count is still below the threshold, we keep the index of that clause and continue. Otherwise, we return the clause to the formula and continue to the next clause. We return a set of the indices of the removed clauses.


We use \ourtool's add and remove clause framework to interactively remove clauses according to the soft core algorithm.
We use a threshold of $\delta=20$ percent for all our tests.
An example for a typical run in this mode is giving a CNF formula as an input to \ourtool 
for our baseline soft core application (described in RQ3).
The tool would modify that CNF {\tt \#Clauses} times and run model counting for each s.t. $\form _1$ is the original CNF and $\form _i$ is the modified CNF in the $i$'th iteration of the soft core algorithm. 
Finally, the algorithm outputs the set of clauses comprising the soft core.

We evaluate 3040 formulas across 6 families, including circuit, planning, and competition benchmarks (details in Appendix~\ref{app:scFamilies}). 
A benchmark is considered solved only if all its constituent instances are processed without timeout.
We excluded from our comparisons benchmarks that timed out on all the tools. 
The families are also classified into 5 runtime intervals, according to the runtime of solving $\form_1$ with $d4$. This was done to better observe the difference between benchmarks in which already the first CNF instance is easier/harder to solve. 


\begin{table*}[t]
\centering
\caption{Best solver for each Soft Core benchmark family by runtime category}
\label{tab:softcore-solver-comparison}
\footnotesize
\begin{tabular}{|l|c|c|c|c|c|}
\hline
\textbf{Family} & \textbf{0--1s} & \textbf{1--10s} & \textbf{10--100s} & \textbf{100--1000s} & \textbf{1000--10000s} \\
\hline
BMS (500) & -- & \shortstack{S (223/242) \\ {\color{brown}G:0/242}} & \shortstack{S (238/254) \\ {\color{brown}G:0/254}} & \shortstack{SS (3/4) \\ {\color{brown}G:0/4}} & -- \\
\hline
CBS (1000) & -- & \shortstack{S (344/344) \\ {\color{brown}G:0/344}} & \shortstack{S (646/656) \\ {\color{brown}G:0/656}} & -- & -- \\
\hline
mcCompetition (113) & \shortstack{SS+TD (16/50) \\ {\color{brown}G:18/50}} & \shortstack{NS+TD (24/34) \\ {\color{brown}G:2/34}} & \shortstack{SS (10/27) \\ {\color{brown}G:2/27}} & \shortstack{SS+TD (1/1) \\ {\color{brown}G:0/1}} & \shortstack{SS+TD (1/1) \\ {\color{brown}G:0/1}}  \\
\hline
or (434) & \shortstack{SS (189/193) \\ {\color{brown}G:0/193}} & \shortstack{SS (83/108) \\ {\color{brown}G:1/108}} & {\color{black} \shortstack{SS+TD (11/61) \\ {\color{brown}*G(42/61)}}} & {\color{black}\shortstack{SS+TD (2/72) \\ {\color{brown}*G(70/72)}}} & -- \\
\hline
planning (616) & \shortstack{SS (133/199) \\ {\color{brown}G:2/199}} & \shortstack{SS (41/120) \\ {\color{brown}G:0/120}} & \shortstack{S+TD (35/124) \\ {\color{brown}G:1/124}} & \shortstack{NS+TD (21/83) \\ {\color{brown}G:1/83}} & \shortstack{S+TD (60/90) \\ {\color{brown}G:6/90}}  \\
\hline
rnd3sat (377) & \shortstack{SS (234/284) \\ {\color{brown}G:0/284}} & \shortstack{SS (88/89) \\ {\color{brown}G:0/89}} & \shortstack{SS (4/4) \\ {\color{brown}G:0/4}} & -- & -- \\
\hline
\end{tabular}

~\\
Legend: NS = NoShared, S = Shared, SS = Shared+Sym, +TD = with Tree Decomposition. G = \ganak.
\\
\emph{Tool (X/Y)} indicates  that Tool outperformed the others tools in X out of Y benchmarks in that cell.

\vspace{-1.5em}


\end{table*}


\subsubsection{Evaluation Details\label{sec:softcore_discussion}}



\subsubsection{Answering RQ1} Consistent with the argumentation results, DLCS outperforms VSADS by comparable margins. Consequently, we employ DLCS as the default branching heuristic for the remainder of our evaluation.

\subsubsection{Answering RQ2} Table~\ref{tab:softcore-solver-comparison} aggregates the results for \ourtool across its three caching modes, evaluated both with and without shared tree decompositions (six configurations in total), plus the external tools {\tt PBCount2} and \ganak.
To compare against  {\tt PBCount2},
we translated the CNFs of our experiments to OPB format that {\tt PBCount2} accepts as input and created an input script of add and remove clause commands to input to {\tt PBCount2}'s incremental mode, in order to emulate our soft core algorithm. Our experiments showed that {\tt PBCount2} timed-out on almost all benchmarks from the 0-1s and 1-10s interval categories, leading us to conclude that its usage of knowledge compilation based model counting may not be suitable for the purpose of our experimental setups.
In order to evaluate our approach with \ganak (in default mode with no extra parameters), we created a soft core tool in C++, based on our own soft core algorithm, that calls \ganak as a black box for the model counting tasks. Since these system calls are being run in a loop sequentially, they cannot share cache or any other information with each other, as opposed to our tool.

\vspace{2mm}

Table~\ref{tab:softcore-solver-comparison} should be read as follows. Each line in the table describes a family of benchmarks with the number of benchmarks that completed the computation (i.e. did not time-out) for at least one tool. 
We checked in how many benchmarks each tool outperformed the others, in terms of runtime. Each cell in the table has in top (black), the  \ourtool mode that won the most benchmarks in its category (family + time interval), while the bottom (brown) shows the number of times that \ganak outperformed the other tools. The cases in which \ganak won overall more benchmarks than any mode of \ourtool are marked with (*).

We first discuss the comparison of our own modes. 
Table~\ref{tab:softcore-solver-comparison} clearly demonstrates that {\tt Shared+Sym} outperforms {\tt NoShared} on easier benchmarks, solving more instances across families such as \emph{BMS}, \emph{CBS}, \emph{rnd3sat}, \emph{or}, and \emph{planning} (within a 10-second runtime threshold).
Similarly to AFs, the lazy symmetry mode added an extra minor boost to the runtime improvement, but the major boost came from using shared cache across benchmarks.
On more challenging benchmarks, such as those from model counting competition and the harder intervals for \emph{planning} and \emph{or} families, Tree decomposition sharing becomes increasingly beneficial. 
For simpler benchmarks, the initial overhead of computing a tree decomposition may exceed the potential search time savings.
However, for harder benchmarks this investment pays off; the TD-enabled modes successfully completed several runs where non-TD modes timed out.
Among the TD configurations, {\tt Shared} modes achieved improvements up to 50\%, though they occasionally favored the \texttt{NS+TD} configuration in specific categories (see Appendix~\ref{app:scComparison} for detailed comparisons). 
Mirroring our AF results, these results suggest that while cache sharing is highly effective in easier settings, its utility diminishes as complexity increases, at which point shared structural information like tree decompositions becomes the primary driver of efficiency.

\begin{figure}[b!]
    \centering
    \includegraphics[width=0.9\columnwidth]{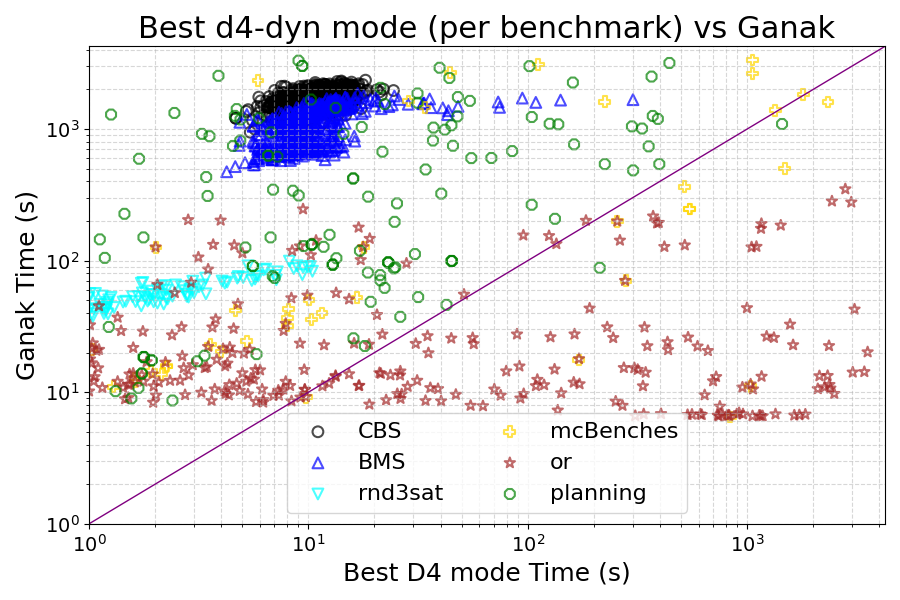}
    \caption{\label{fig:best_d4_vs_ganak} Runtimes of \ourtool best mode vs \ganak}
\end{figure}

\paragraph{ Answering RQ3}
We next compare \ourtool{} against {\tt ganak}.
As shown in Table~\ref{tab:softcore-solver-comparison}, \ourtool{} exhibits superior performance on easier benchmarks.
This advantage stems from our integrated framework, which facilitates knowledge sharing and avoids the repeated process initialization overhead inherent in \ganak's baseline execution.
On the harder benchmarks on which \ganak is known to be better \cite{DBLP:journals/corr/abs-2504-13842} it shines, especially since, as discussed before, there is less utility in sharing on the harder benchmarks. 
Specifically, on harder benchmarks, {\tt ganak}'s specialized search techniques allow it to excel, particularly as the marginal utility of cache sharing diminishes with increasing problem complexity.
Notably, \ganak exhibits exceptional performance on the \emph{or} benchmarks, where its robust pre-processing is often sufficient to solve instances entirely without further search.
This is further illustrated in Figure~\ref{fig:best_d4_vs_ganak}, which compares the runtime of \ourtool's optimal configuration against that of \ganak across all benchmarks. 
These results corroborate Table~\ref{tab:softcore-solver-comparison}, demonstrating that \ourtool outperforms \ganak on the majority of benchmark families, with the \emph{or} benchmarks remaining a notable exception, where \ganak's efficiency is most pronounced.

All in all, our results demonstrate the consistent advantage of knowledge reuse across nearly all benchmarks. While shared caching and symmetry often suffice to improve performance, the reuse of tree decompositions proves particularly effective for the most challenging benchmarks.


\section{Conclusion}\label{sec:conclusion}

In this work, we tailored an incremental model counter, \ourtool, that rather than treating each benchmark independently, re-uses both cache information and branching-heuristic knowledge 
between benchmarks to exploit their similarities. We evaluated our approach on dynamic argumentation benchmarks, where syntactic similarity is crucial, for which we showed how our sharing techniques improve over existing solvers. 
We also considered the computation of minimal subset soft cores, showing that our framework can leverage cache and heuristic reuse to improve efficiency in this setting, and outperforms existing tools using benchmarks taken from competitions. 
Our results highlight the benefits of cache reuse within our benchmark sequences.
To the best of our knowledge, this is the first work to exploit cache and heuristic reuse across multiple benchmarks in DPLL-based model counting. 
As this paper introduces a novel setting, it serves as a foundational first step. In both problems that we chose, model counting and the notion of incrementality is natural. Showing that our framework can handle further problems is a promising future work that we intend on pursuing.

\section*{Acknowledgements}
This work has been partly supported by the CERADOC project of the French National Agency for Research (ANR-25-CE23-3078).
We thank Mate Soos for his advice during the early stages of this work.
We thank the reviewers for their insights and suggestions that helped improve this paper.



\bibliographystyle{kr}
\bibliography{mybibfile}

\clearpage

\appendix

\section{Supplementary Material}

\subsection{Families of Argumentation Frameworks}\label{app:afFamilies}

We describe the families of benchmarks that we used in our Argumentation Framework experiments.
We use 263 CNF repositories from the 2023 Abstract Argumentation Competition, grouped into 14 datasets, for each family of repositories we list the minimal, maximal and average size of the argumentation frameworks in that repository.

\begin{table}[h]
\centering
\scalebox{1}{
\begin{tabular}{lcccc}
\toprule
\textbf{Repository} & \textbf{\#Benchmarks} & \textbf{Min} & \textbf{Max} & \textbf{Average} \\
\midrule
\emph{admbuster}   & 10  & 1000  & 200161 & 40170.95 \\
\emph{afinput}     & 25  & 108   & 1517   & 418.50 \\
\emph{bw}          & 2   & 108   & 1517   & 418.50 \\
\emph{BA}          & 25  & 101   & 339    & 220.70 \\
\emph{cities}      & 25  & 116   & 8079   & 446.20 \\
\emph{ER}          & 25  & 101   & 453    & 262.21 \\
\emph{g}           & 25  & 1160  & 3821   & 2373.85 \\
\emph{medium}      & 19  & 2914  & 2948   & 2931.73 \\
\emph{n}           & 15  & 100   & 405    & 235.25 \\
\emph{scc}         & 25  & 226   & 1963   & 1226.60 \\
\emph{sembuster}   & 15  & 150   & 1643   & 847.78 \\
\emph{small}       & 2   & 191   & 338    & 267.18 \\
\emph{st}          & 25  & 319   & 585    & 465.57 \\
\emph{WS}          & 25  & 100   & 644    & 298.95 \\
\bottomrule
\end{tabular}}

\caption{Statistics of the argumentation framework repositories.}
\label{tab:af-stats}
\end{table}

\subsection{Additional Soft Core Comparisons}\label{app:scComparison}

We describe the comparison that we made between the different modes of our tool.

\begin{figure}[htb!]
    \centering
    \includegraphics[width=0.9\columnwidth]{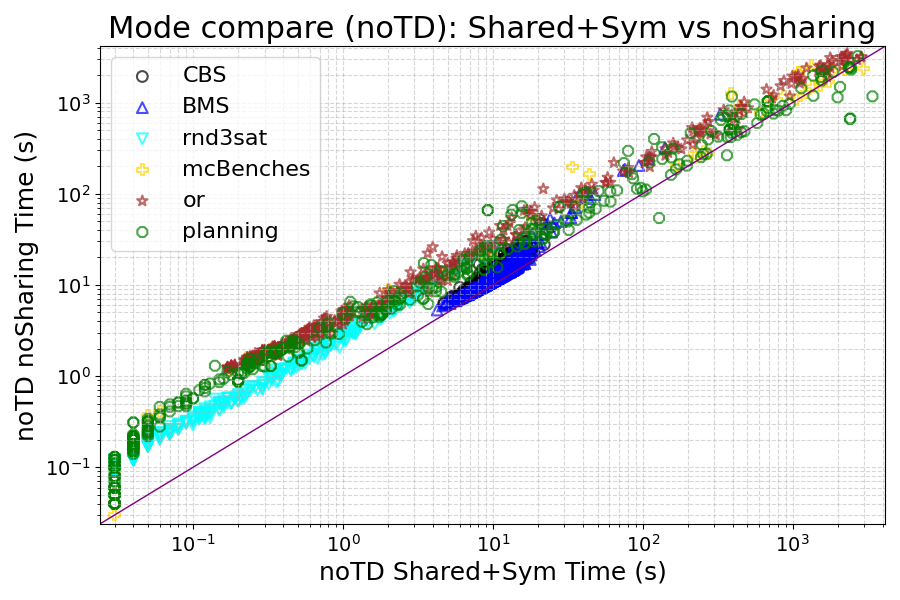}
    \caption{\label{fig:noTD_sharedSym_vs_nosharing} runtimes of \ourtool({\tt NoShared}) and \ourtool({\tt Shared+Sym}) modes without TD sharing.}
\end{figure}

Figure~\ref{fig:noTD_sharedSym_vs_nosharing} Compares the run times of \ourtool({\tt NoShared}) and \ourtool({\tt Shared+Sym}) modes without TD sharing. As can be seen, plot highlights the performance gain of using Shared+Sym, especially on easier benchmarks.

\begin{figure}[htb!]
    \centering
    \includegraphics[width=0.9\columnwidth]{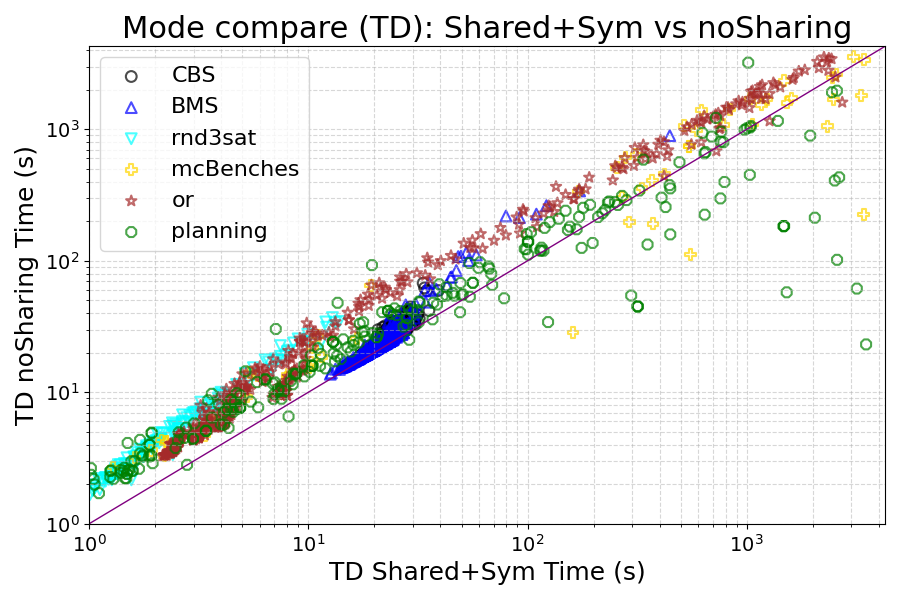}
    \caption{\label{fig:TD_sharedSym_vs_nosharing} runtimes of \ourtool({\tt NoShared} and \ourtool({\tt Shared+Sym}) modes with TD sharing.}
\end{figure}

Figure~\ref{fig:TD_sharedSym_vs_nosharing} compares the run times of \ourtool({\tt NoShared} and \ourtool({\tt Shared+Sym}) modes with TD sharing. As can be seen, this plot highlights the performance gain of using TD, but the gain from Shared+Sym is more moderate.

\begin{figure}[h]
    \centering
    \includegraphics[width=0.9\columnwidth]{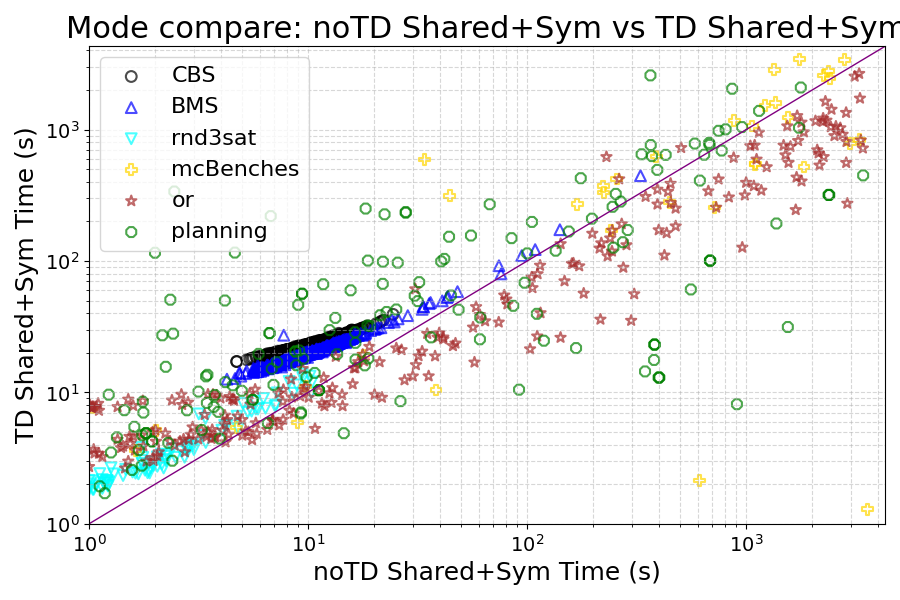}
    \caption{\label{fig:shared_sym_td_no_td} runtimes of \ourtool({\tt Shared+Sym}, {\tt DLCS}) modes and without without TD sharing.}
\end{figure}

Finally, Figure~\ref{fig:shared_sym_td_no_td} Compares the run times of \ourtool({\tt Shared+Sym}, {\tt DLCS}) modes and without without TD sharing. As can be seen, this plot highlights the gain of using TD sharing on harder benchmarks, and that it is better not to use it on easier benchmarks.

\subsection{Families of Soft core Benchmarks}\label{app:scFamilies}

Table~\ref{tab:softcore_datasets} below describes our soft core benchmarks, categorized by families. Each line describes the family name, number of benchmarks, number of benchmarks that had a successful run on at least one tool (\#w/oT TOs), and a description of the family.

\begin{table}[htb!]
\centering
\renewcommand{\arraystretch}{1.2}
\begin{tabular}{l c c p{12cm}}
\toprule
Dataset & \#  & \# w/o TOs & Description \\
\midrule
BMS & 500 & 500 & Benchmarks from the Boolean Model Counting Suite. Considered easier. \\
CBS & 1000 & 1000 & Circuit-based SAT instances, often derived from hardware verification problems. Considered easier. \\
mcBenches & 784 & 113 & Benchmarks taken from the model counting competitions of 2021--2024 designed to stress-test solver performance. Considered harder. \\
or & 660 & 434 & Operations research benchmark family. Considered harder when not using preprocessing but easier with preprocessing enabled. \\
planning & 951 & 616 & Miscellaneous SAT instances from various domains such as Random synthetic SATLIB constructions (Schur, Skolem, Ramsey), Blasted, Scenarios, Countdump, NoCountDump, Circuit-derived Benchmarks, PDDL, qg, Controller verification Benchmarks, Empty Room, Blocks world, Ring navigation, UTS, Logistics, Sorting/Numbering tasks, Flip/No-Action Planning. \\
rnd3sat & 410 & 377 & Randomly generated 3-SAT formulas. Considered easy to mid-range. \\
\bottomrule
\end{tabular}
\caption{Overview of the soft core benchmarks}
\label{tab:softcore_datasets}
\end{table}

\clearpage

\end{document}